\documentclass[12pt]{article}
\usepackage{amsmath}
\usepackage{amssymb}
\usepackage{amsfonts}

\begin{document}

\title{ Disconjugacy, regularity of multi-indexed rationally-extended
potentials, and Laguerre exceptional polynomials}
\author{Y. Grandati$^{1}$ and C. Quesne$^{2}$ \\
{\small \textsl{$^1$Equipe BioPhysStat, LCP A2MC, Universit\'e de
Lorraine-Site de Metz,}}\\
{\small \textsl{1 bvd D. F. Arago, F-57070, Metz, France}}\\
{\small \textsl{grandati@yahoo.fr}}\\
{\small \textsl{$^2$Physique Nucl\'eaire Th\'eorique et Physique
Math\'ematique, Universit\'e Libre de Bruxelles,}} \\
{\small \textsl{Campus de la Plaine CP229, Boulevard~du Triomphe, B-1050
Brussels, Belgium}}\\
{\small \textsl{cquesne@ulb.ac.be}}}
\date{ }
\maketitle

\begin{abstract}
The power of the disconjugacy properties of second-order differential
equations of Schr\"odinger type to check the regularity of
rationally-extended quantum potentials connected with exceptional orthogonal
polynomials is illustrated by re-examining the extensions of the isotonic
oscillator (or radial oscillator) potential derived in $k$th-order
supersymmetric quantum mechanics or multistep Darboux-B\"acklund
transformation method. The function arising in the potential denominator is
proved to be a polynomial with a nonvanishing constant term, whose value is
calculated by induction over $k$. The sign of this term being the same as
that of the already known highest-degree term, the potential denominator has
the same sign at both extremities of the definition interval, a property
that is shared by the seed eigenfunction used in the potential construction.
By virtue of disconjugacy, such a property implies the nodeless character of
both the eigenfunction and the resulting potential.
\end{abstract}

\baselineskip=22pt plus 1pt minus 1pt 


%
\newpage

\section{INTRODUCTION}

The introduction of the concept of exceptional orthogonal polynomials (EOP) 
\cite{gomez04a, gomez04b} and the construction of the first $X_1$ families
of Laguerre- and Jacobi-type EOP in the context of Sturm-Liouville theory 
\cite{gomez09, gomez10a} have aroused a lot of interest both in mathematics
and in physics. On the one hand, the EOP families are novel systems of
complete orthogonal polynomials with respect to some positive-definite
measure, generalizing the classical families of Hermite, Laguerre, and
Jacobi, but differing from these among others by the presence of gaps in the
degrees $n$ of the polynomials appearing in the sequences (e.g., $n=0$ for
the $X_1$ families). On the other hand, they were shown \cite{cq08} to
provide a keystone for rationally extending some well-known
(translationally) shape-invariant quantum potentials \cite{gendenshtein,
cooper, carinena00} in such a way that shape invariance is preserved. In
this context, the usefulness of an approach based on supersymmetric quantum
mechanics (SUSYQM) or, equivalently, the Darboux transformation was pointed
out \cite{bagchi, cq09}.

%
%
An important step was the description of infinitely many shape-invariant
potentials associated with $X_m$ EOP families, where the number of gaps $m$,
called codimension (and corresponding here to $n=0$, 1, \ldots,~$m-1$), may
be arbitrarily large \cite{odake09, odake10a, odake10b, ho11a}. The
existence of two families of $X_m$-Laguerre and $X_m$-Jacobi EOP was also
reported (thereby extending an observation made for $m=2$ in Ref.\ \cite%
{cq09}) and it was later on explained through Darboux-Crum transformation 
\cite{sasaki, gomez10b, gomez12a}. In addition, these families were shown to
be obtainable through two alternative approaches, namely the
Darboux-B\"acklund transformation \cite{grandati11a} and the prepotential
method \cite{ho11b}.

%
%
Counterparts of the classical family of Hermite polynomials have also been
known in physical applications since the early 90s \cite{dubov92, dubov94,
bagrov, samsonov, junker97, junker98, carinena08, fellows}, but these $X_m$
EOP are of a slightly different nature in the sense that the $m$ gaps in the
degree sequence correspond to $n=1$, 2, \ldots,~$m$, instead of $n=0$, 1,
\ldots,~$m-1$. Since then, similar $X_m$-Laguerre and $X_m$-Jacobi EOP have
been constructed and form the so-called third families \cite{cq09,
grandati11a, ho11b}.

%
%
A significant advance was then the introduction of multi-indexed $X_{m_1,
m_2, \ldots, m_k}$ EOP families and associated rationally-extended
potentials through multistep Darboux algebraic transformations \cite%
{gomez12b}. The same concept was developed through several approaches, such
as the Crum-Adler mechanism \cite{odake11}, higher-order SUSYQM \cite{cq11a,
cq11b}, and multistep Darboux-B\"acklund transformations \cite{grandati12}.

%
%
In such a context, there remain two open essential questions. The first one
is to know whether the multi-indexed EOP families exhaust all the
possibilities of higher-codimensional complete orthogonal polynomial systems
or, in other words, whether all the higher-codimensional complete orthogonal
polynomial systems are generated by the application of successive algebraic
Darboux transformations (or one of the equivalent methods quoted above).
This was recently proved to be true for codimension two \cite{gomez12c}.

%
%
The second problem has to do with the existence of a well-behaved measure
defining an EOP system, which is directly connected with the regularity of
the associated rationally-extended quantum potentials. In the one-indexed $%
X_m$ case, this problem is usually solved by showing that the classical
polynomial occurring in the denominator has no zero in the domain of the
variable, which is an easy task since the distribution of zeros of classical
polynomials is well known \cite{erdelyi, szego}. In the multi-indexed $%
X_{m_1, m_2, \ldots, m_k}$ case, however, the denominator contains a
Wronskian of functions written in terms of classical polynomials, so that
the problem becomes much more difficult to solve. While still feasible for $%
k=2$, the study of zeros remains outside scope for most higher $k$ cases,
hence the construction of multi-indexed EOP is still rather formal.

%
%
A notable progress towards the solution of the regularity question was
recently made \cite{grandati12} by making use of the disconjugacy properties
of the Schr\"{o}dinger equation for eigenvalues below the ground state \cite%
{hartman, coppel, bocher}. As a consequence of such properties, the nodeless
character of the function present in the denominator can be inferred from
the signs it takes on both ends of the domain. Nevertheless when applied to
the rationally-extended isotonic oscillator (or radial oscillator)
potentials \cite{grandati12}, except in the $k=2$ case, the equality of
signs was obtained without exhibiting explicit formulas for the denominator
in the asymptotic regions.

%
%
The purpose of the present paper is to give such explicit expressions in the
general multi-indexed case. We then extend some results previously obtained
in \cite{cq11a, cq11b}, confirming in particular a conjecture about the
structure of the polynomial factor in the denominator. This allows a more
direct proof of the regular character of the considered extensions.

%
%
The paper is organized as follows. After recalling in Sec.~II the
disconjugacy properties of the Schr\"{o}dinger equation, we apply it to the
construction of rationally-extended radial oscillator potentials and
corresponding $X_{m}$-Laguerre EOP in first-order SUSYQM. Section III
summarizes the corresponding multistep construction in $k$th-order SUSYQM,
where each step may be associated with any one of the first two families of $%
X_{m}$-Laguerre EOP. An improved proof of the regularity of these extensions
is then given in Sec.~IV. Finally, Section V contains the conclusion.

%
%

\section{\boldmath RATIONAL EXTENSIONS OF THE ISOTONIC POTENTIAL AND $X_m$%
-LAGUERRE EOP}

\subsection{Disconjugacy}

Let us briefly recall some essential elements about the disconjugacy
properties of second-order differential equations of Schr\"odinger type, 
\begin{equation}
\psi^{\prime \prime }(x) + \bigl(E - V(x)\bigr) \psi(x) = 0.  \label{eq:EdS}
\end{equation}

%
%
To any solution $\psi (x)$ of Eq.\ (\ref{eq:EdS}) we can associate the
corresponding Ricatti-Schr\"{o}dinger function $w(x)=-\psi ^{\prime
}(x)/\psi (x)$, which is a solution of 
\begin{equation}
-w^{\prime }(x)+w^{2}(x)=V(x)-E.  \label{eq:RS}
\end{equation}

%
%
Equation (\ref{eq:EdS}) is said to be disconjugated on $I \subset \mathbb{R}$
($V(x)$ being supposed continuous on $I$) if every solution of this equation
has at most one (necessarily simple) zero on $I$ \cite{hartman, coppel}. For
a closed or open interval $I$, the disconjugacy of Eq.\ (\ref{eq:EdS}) is
equivalent to the existence of solutions that are everywhere non zero on $I$ 
\cite{hartman, coppel}. In the following we will consider $I = ]0, +\infty[$.

%
%
Let us also quote the following result: \newline
\textbf{Theorem} \cite{hartman, coppel} \textit{If there exists a
continuously differentiable solution on $I$ of the Riccati inequality} 
\begin{equation}
-w^{\prime }(x)+w^{2}(x)\leq V(x)-E,  \label{eq:Ricineq}
\end{equation}%
\textit{then Eq.\ (\ref{eq:EdS}) is disconjugated on $I$.}

%
%
In the case we consider, the Hamiltonian corresponding to Eq.\ (\ref{eq:EdS}%
) and associated to Dirichlet boundary conditions on $I$ has an infinite
bound-state spectrum $\{ E_{\nu}, \psi_{\nu}\}_{\nu \in \mathbb{N}}$.
Consequently, since $\psi_0$ is nodeless on $I$, $w_0(x)$ is a continuously
differentiable solution of (\ref{eq:RS}) on this domain and condition (\ref%
{eq:Ricineq}) is fulfilled for every value of $E$ not greater than $E_0$.
The domain of values $]-\infty, E_0]$ of the spectral parameter $E$ will be
called the disconjugacy sector of Eq.\ (\ref{eq:EdS}). In this sector,
except for $E=E_0$, $\psi(x)$ does not satisfy the required Dirichlet
boundary conditions to be a bound-state wavefunction and diverges at one or
both extremities of $I$. To ensure that this eigenfunction is nodeless, it
is sufficient to verify that it has the same sign at both extremities of the
definition interval.

%
%

\subsection{\boldmath Isotonic oscillator and $X_m$-Laguerre EOP}

The isotonic oscillator potential is defined on the positive half-line $x >
0 $ as 
\begin{equation}
V_l(x) = \frac{1}{4} \omega^2 x^2 + \frac{l(l+1)}{x^2}.  \label{eq:potiso}
\end{equation}

%
%
The spectrum of the associated Schr\"odinger equation has an infinite number
of bound states, whose (unnormalized) wavefunctions are given by 
\begin{equation}
\psi^{(l)}_{\nu}(x) \propto \eta_l(z) L^{(\alpha)}_{\nu}(z), \qquad \nu \in 
\mathbb{N},
\end{equation}
where $z = \omega x^2/2$, $\alpha = l + 1/2$, and 
\begin{equation}
\eta_l(z) = z^{(\alpha + 1/2)/2} e^{-z/2},
\end{equation}
$L^{(\alpha)}_{\nu}(z)$ being a generalized Laguerre polynomial \cite%
{erdelyi}. The associated energies are 
\begin{equation}
E^{(l)}_{\nu} = \omega (2\nu + \alpha + 1).
\end{equation}

%
%
In first-order SUSYQM, the superpartner Hamiltonians are defined as \cite%
{cooper} 
\begin{equation}
\left\{ 
\begin{array}{c}
H^{(+)}=A^{\dagger }A=-\frac{\textstyle{d^{2}}}{\textstyle{dx^{2}}}%
+V^{(+)}(x)-E, \\[0.2cm]
H^{(-)}=AA^{\dagger }=-\frac{\textstyle{d^{2}}}{\textstyle{dx^{2}}}%
+V^{(-)}(x)-E,%
\end{array}%
\right. 
\end{equation}%
with 
\begin{equation}
\left\{ 
\begin{array}{c}
A^{\dagger }=-{\frac{\textstyle{d}}{\textstyle{dx}}}+W(x), \\[0.2cm]
A={\frac{\textstyle{d}}{\textstyle{dx}}}+W(x),%
\end{array}%
\right. 
\end{equation}%
where the superpotential $W(x)$ satisfies the Riccati equations 
\begin{equation}
\mp W^{\prime }(x)+W^{2}(x)=V^{(\pm )}(x)-E.
\end{equation}

%
%
$W(x)$ is obtained from a solution $\phi(x)$ of the initial Schr\"odinger
equation 
\begin{equation}
H^{(+)} \phi(x) = 0,
\end{equation}
via 
\begin{equation}
W(x) = - \phi^{\prime }(x)/\phi(x).
\end{equation}

%
%
The new potential $V^{(-)}(x) = V^{(+)}(x) + 2W^{\prime }(x)$ is regular as
soon as the ``seed'' solution $\phi(x)$ is nodeless on $]0, +\infty[$. This
can be achieved if we choose the factorization energy in the disconjugacy
sector of $H^{(+)}$, that is, when $E$ is not greater than the ground-state
energy $E^{(+)}_0$ of $V^{(+)}(x)$. This is true in particular if we choose
as seed solution the ground-state wavefunction $\psi^{(l)}_0(x) = \eta_l(z)$
itself. For $E < E^{(+)}_0$, as discussed above, the analysis of the sign of 
$\phi(x)$ at $0^+$ and $+\infty$ is then sufficient to control the
regularity of $W(x)$.

%
%
Suppose that we are in this case. Then we have three possibilities (up to a
global change of sign), namely 
\begin{equation}
\left\{ 
\begin{array}{c}
\mathrm{I}: \phi(0^+) = 0^+, \quad \phi(+\infty) = + \infty, \\[0.2cm] 
\mathrm{II}: \phi(0^+) = + \infty, \quad \phi(+\infty) = 0^+, \\[0.2cm] 
\mathrm{III}: \phi(0^+) = + \infty, \quad \phi(+\infty) = + \infty.%
\end{array}
\right.  \label{eq:BCseed}
\end{equation}

%
%
In the first two cases, $1/\phi(x)$ diverges at one extremity of the
positive half-line and, although we have 
\begin{equation}
H^{(-)} \left(\frac{1}{\phi(x)}\right) = 0,  \label{eq:fondpartner}
\end{equation}
it is not a bound-state wavefunction of $H^{(-)}$. Then $H^{(+)}$ and $%
H^{(-)}$ turn out to be isospectral. If we are looking for seed functions
which are polynomials (up to some gauge factor), we obtain them from the
eigenstates by applying the following discrete symmetries acting on the
parameters of the initial potential (\ref{eq:potiso}) \cite{grandati11a} 
\begin{equation}
\left\{ 
\begin{array}{c}
\Gamma_{\mathrm{I}}: (\omega, \alpha) \to (-\omega, \alpha), \\[0.2cm] 
\Gamma_{\mathrm{II}}: (\omega, \alpha) \to (\omega, -\alpha), \\[0.2cm] 
\Gamma_{\mathrm{III}}: (\omega, \alpha) \to (-\omega, -\alpha).%
\end{array}
\right.  \label{eq:sym}
\end{equation}

%
%
By applying the symmetry $\Gamma_i$, $i=$ I, II, III, to $\psi^{(l)}_{\nu}$,
we obtain a seed eigenfunction $\phi_{\nu}$, which satisfies boundary
conditions of type $i$ (see (\ref{eq:BCseed})). In particular, for $i=$ I or
II, we obtain 
\begin{equation}
\left\{ 
\begin{array}{c}
\Gamma_{\mathrm{I}}: \psi^{(l)}_m(x) \to \phi^{\mathrm{I}}_{lm}(x) = \chi^{%
\mathrm{I}}_l(z) L^{(\alpha)}_m(-z), \\[0.2cm] 
\Gamma_{\mathrm{II}}: \psi^{(l)}_m(x) \to \phi^{\mathrm{II}}_{lm}(x) = \chi^{%
\mathrm{II}}_l(z) L^{(-\alpha)}_m(z),%
\end{array}
\right.  \label{eq:seeds}
\end{equation}
where 
\begin{equation}
\left\{ 
\begin{array}{c}
\chi^{\mathrm{I}}_l(z) = z^{(\alpha+1/2)/2} e^{z/2}, \\[0.2cm] 
\chi^{\mathrm{II}}_l(z) = z^{-(\alpha-1/2)/2} e^{-z/2},%
\end{array}
\right.  \label{eq:gauges}
\end{equation}
for the corresponding energies 
\begin{equation}
\left\{ 
\begin{array}{c}
E^{\mathrm{I}}_{lm} = - \omega (\alpha + 2m + 1), \\[0.2cm] 
E^{\mathrm{II}}_{lm} = - \omega (\alpha - 2m - 1).%
\end{array}
\right.  \label{eq:negenergies}
\end{equation}

%
%
If $\phi^{\mathrm{I}}_{lm}$ is always in the disconjugacy sector, for $\phi^{%
\mathrm{II}}_{lm}$ to reach it we need to assume the condition $\alpha > m$.
Starting from the isotonic potential $V^{(+)}(x) = V_{l^{\prime }}(x)$, the
SUSYQM partnership based on seed eigenfunctions $\phi^{\mathrm{I}%
}_{l^{\prime }m}$ or $\phi^{\mathrm{II}}_{l^{\prime }m}$ with respectively $%
l^{\prime }= l-1$ and $l^{\prime }= l+1$ gives then rational extensions of
the form \cite{cq11b} 
\begin{equation}
V^{(-)}(x) = V_l(x) + V_{l,\mathrm{rat}}(x) + C,
\end{equation}
where 
\begin{equation}
V_{l,\mathrm{rat}}(x) = - 2\omega \left\{\frac{\dot{g}^{(\alpha)}_m}{%
g^{(\alpha)}_m} + 2z \left[\frac{\ddot{g}^{(\alpha)}_m}{g^{(\alpha)}_m} -
\left(\frac{\dot{g}^{\alpha)}_m} {g^{(\alpha)}_m}\right)^2\right]\right\}
\end{equation}
(the dot denoting a derivative with respect to $z$), with in the first case 
\begin{equation}
g^{(\alpha)}_m(z) = L^{(\alpha-1)}_m(-z), \qquad C = - \omega
\end{equation}
and in the second case 
\begin{equation}
g^{(\alpha)}_m(z) = L^{(-\alpha-1)}_m(z), \qquad C = \omega.
\end{equation}

%
%
Due to disconjugacy properties, the regularity of these extensions on the
positive half-line is ensured for every $m=1$, 2,~\ldots\ (and $\alpha$
large enough for type II).

%
%
The bound-state wavefunctions of the superpartner potential $V^{(-)}(x)$ are
obtained by applying the operator $A$ (with $W(x) = - \phi^{i
\prime}_{l^{\prime }m}(x)/\phi^{i}_{l^{\prime }m}(x)$, $i=$ I, II) on those
of $V^{(+)}(x)$ ($\psi^{(+)}_{\nu}(x) \propto \eta_{l^{\prime }}(z)
L^{(\alpha^{\prime })}_{\nu}(z)$ with $\alpha^{\prime }= l^{\prime }+ 1/2$).
We can write 
\begin{equation}
\psi^{(-)}_{\nu}(x) \propto \frac{\eta_l(z)}{g^{(\alpha)}_m(z)}
y^{(\alpha)}_{m+\nu}(z), \qquad \nu=0, 1, 2, \ldots,
\end{equation}
where $y^{(\alpha)}_{m+\nu}(z)$ is a polynomial of degree $n = m+\nu$, 
\begin{equation}
y^{(\alpha)}_n(z) = \left\{ 
\begin{array}{ll}
L^{\mathrm{I}}_{\alpha,m,n}(z) & \mathrm{if\ } i = \mathrm{I}, \\[0.2cm] 
L^{\mathrm{II}}_{\alpha,m,n}(z) & \mathrm{if\ } i = \mathrm{II},%
\end{array}
\right.
\end{equation}
satisfying the second-order differential equation 
\begin{equation}
\left[z \frac{d^2}{dz^2} + \left(\alpha+1-z - 2z \frac{\dot{g}^{(\alpha)}_m}{%
g^{(\alpha)}_m}\right) \frac{d}{dz} + (z-\alpha) \frac{\dot{g}^{(\alpha)}_m}{%
g^{(\alpha)}_m} + z \frac{\ddot{g}^{(\alpha)}_m}{g^{(\alpha)}_m}\right]
y^{(\alpha)}_{m+\nu}(z) = - \nu y^{(\alpha)}_{m+\nu}(z).
\end{equation}

%
%
The set of polynomials $y^{(\alpha)}_{m+\nu}(z)$ ($\nu= 0$, 1, 2,~\ldots),
defined on the positive half-line, is orthogonal and complete with respect
to the positive-definite measure of density 
\begin{equation}
z^{\alpha} e^{-z}/\bigl( g^{(\alpha)}_m(z)\bigr)^2.
\end{equation}

%
%

\section{\boldmath LAGUERRE EOP AND REDUCIBLE $k$-TH ORDER SUSYQM}

In $k$th-order SUSYQM, the first-order differential operators $A$ and $%
A^{\dagger}$ are replaced by $k$th-order ones $\mathcal{A}$ and $\mathcal{A}%
^{\dagger}$ \cite{bagrov, samsonov, andrianov93, andrianov95, fernandez}.
The correspondence between the initial and final Hamiltonians $H^{(1)}$ and $%
H^{(2)}$ is determined by the intertwining relations 
\begin{equation}
\mathcal{A} H^{(1)} = H^{(2)} \mathcal{A}, \qquad \mathcal{A}^{\dagger}
H^{(2)} = H^{(1)} \mathcal{A}^{\dagger}.
\end{equation}

%
%
In the reducible case, the operators $\mathcal{A}$ and $\mathcal{A}%
^{\dagger} $ can be factorized into products of first-order differential
operators as 
\begin{equation}
\mathcal{A} = A^{(k)} A^{(k-1)} \cdots A^{(1)},  \label{eq:An}
\end{equation}
where 
\begin{equation}
A^{(i)} = \frac{d}{dx} + W^{(i)}(x), \qquad W^{(i)}(x) = - \bigl(\log
\phi^{(i)}(x)\bigr)^{\prime },
\end{equation}
the seed eigenfunctions being given by Crum's formula \cite{crum} 
\begin{equation}
\phi^{(i)}(x) = \phi^{(i)}_{1,2, \ldots, i}(x) = \frac{\mathcal{W}(\phi_1,
\phi_2, \ldots, \phi_i \mid x)} {\mathcal{W}(\phi_1, \phi_2, \ldots,
\phi_{i-1} \mid x)},  \label{eq:phii}
\end{equation}
with $\mathcal{W}(\phi_1, \phi_2, \ldots, \phi_i | x)$ denoting the
Wronskian of $\phi_1(x)$, $\phi_2(x)$, \ldots, $\phi_i(x)$. Here we have
introduced lower indices 1, 2, \ldots, $i$ to specify which seed functions
have been used. The partner potentials are then linked by the second Crum's
formula 
\begin{equation}
V^{(2)}(x) = V^{(1)}(x) - 2 \frac{d^2}{dx^2} \log \mathcal{W}(\phi_1,
\phi_2, \ldots, \phi_k | x).  \label{eq:extpot}
\end{equation}

%
%
As in Sec.~II, we start from the isotonic potential $V^{(1)}(x) =
V_{l^{\prime }}(x)$ to get an extended potential of the form 
\begin{equation}
V^{(2)}(x) = V_l(x) + V_{l,\mathrm{rat}}(x) + C.  \label{eq:extpot-bis}
\end{equation}

%
%
The final potential is of type $\mathrm{I}^q \mathrm{II}^{k-q}$ if the
chosen set of seed eigenfunctions includes $q \le k$ seed functions of type
I and $k-q$ seed functions of type II. The order of the seed functions being
irrelevant, we can assume that 
\begin{equation}
\phi_i(x) = \left\{ 
\begin{array}{ll}
\phi^{\mathrm{I}}_{l^{\prime }m_i}(x) = \chi^{\mathrm{I}}_{l^{\prime }}(z)
L^{(\alpha^{\prime })}_{m_i}(-z), & i=1, \ldots, q, \\[0.2cm] 
\phi^{\mathrm{II}}_{l^{\prime }m_i}(x) =\chi^{\mathrm{II}}_{l^{\prime }}(z)
L^{(-\alpha^{\prime })}_{m_i}(z), & i=q+1, \ldots, k,%
\end{array}
\right.
\end{equation}
with $l^{\prime }= l + k - 2q$ (hence $\alpha^{\prime }= \alpha + k - 2q$), $%
0 < m_1 < \cdots < m_q$, and $0 < m_{q+1} < \cdots < m_k$, these functions
being nodeless for $\alpha^{\prime }> \underset{i=q+1, \ldots, k}{\sup}(m_i)$%
. In \cite{cq11b}, it has been shown that the Wronskian appearing in Crum's
formula (\ref{eq:extpot}) can be written as 
\begin{equation}
\mathcal{W}(\phi_1, \phi_2, \ldots, \phi_k \mid x) = (\omega x)^{k(k-1)/2}
z^{-q(k-q)} \bigl(\chi^{\mathrm{I}}_{l^{\prime }}\bigr)^q \bigl(\chi^{%
\mathrm{II}}_{l^{\prime }}\bigr)^{k-q} g^{(\alpha)}_{\mu}(z),
\label{eq:wronskn}
\end{equation}
where 
\begin{equation}
g^{(\alpha)}_{\mu}(z) = z^{-(k-q)(k-q-1)} \det \tilde{\Gamma}%
^{(\alpha)}_{\mu},  \label{eq:g}
\end{equation}
with 
\begin{equation}
\Bigl(\tilde{\Gamma}^{(\alpha)}_{\mu}\Bigr)_{ij} = \left\{ 
\begin{array}{ll}
L^{(\alpha^{\prime }+ i - 1)}_{m_j - i + 1}(-z), & 1 \le i \le q+1, 1 \le j
\le q, \\[0.2cm] 
(m_j + 1)_{i - 1} z^{k-i} L^{(-\alpha^{\prime }- i + 1)}_{m_j + i - 1}(z), & 
1 \le i \le q+1, q+1 \le j \le k, \\[0.2cm] 
L^{(\alpha^{\prime }+ i - 1)}_{m_j - q}(-z), & q+2 \le i \le k, 1 \le j \le
q, \\[0.2cm] 
(m_j + 1)_q (m_j - \alpha^{\prime }- i + q + 2)_{i-q-1} & {} \\[0.2cm] 
\quad {} \times z^{k-i} L^{(-\alpha^{\prime }- i + 1)}_{m_j + q}(z), & q+2
\le i \le k, q+1 \le j \le k,%
\end{array}
\right.  \label{eq:gamma}
\end{equation}
and $(a)_n = a (a+1) \cdots (a+n-1)$ denoting the usual Pochhammer symbol 
\cite{erdelyi}.

%
%
Observe that on taking (\ref{eq:extpot}) and (\ref{eq:wronskn}) into
account, we can express $V_{l, \mathrm{rat}}(x)$ and $C$ of Eq.\ (\ref%
{eq:extpot-bis}) as 
\begin{equation}
V_{l,\mathrm{rat}}(x) = - 2\omega \left\{\frac{\dot{g}^{(\alpha)}_{\mu}}{%
g^{(\alpha)}_{\mu}} + 2z \left[\frac{\ddot{g}^{(\alpha)}_{\mu}}{%
g^{(\alpha)}_{\mu}} - \left(\frac{\dot{g}^{\alpha)}_{\mu}} {%
g^{(\alpha)}_{\mu}}\right)^2\right]\right\}, \qquad C = (k - 2q) \omega.
\end{equation}

%
%
In \cite{cq11b}, it has been proved that 
\begin{equation}
g^{(\alpha)}_{\mu}(z) = \mathcal{C}^{(\alpha)}_{\mu} z^{\mu} + \text{%
lower-order terms},
\end{equation}
with 
\begin{equation}
\left\{ 
\begin{array}{c}
\mu = \sum_{i=1}^k m_i - q(q-1)/2 - (k-q)(k-q-1)/2 + q(k-q), \\[0.2cm] 
\mathcal{C}^{(\alpha)}_{\mu} = (-1)^{\sigma} \Delta(m_1, \ldots, m_q)
\Delta(m_{q+1}, \ldots, m_k)/(m_1! \ldots m_k!), \\[0.2cm] 
\sigma = \sum_{i=q+1}^k m_i + q(k-q),%
\end{array}
\right.  \label{eq:musigma}
\end{equation}
$\Delta(n_1, \ldots, n_l)$ being a Vandermonde determinant of order $l$, 
\begin{equation}
\Delta(n_1, \ldots, n_l) = \prod_{1\le i<j \le l} (n_j - n_i).
\label{eq:vandermonde}
\end{equation}

%
%
As shown in \cite{grandati12}, the Wronskian (\ref{eq:wronskn}) has no node
on the positive half-line, which ensures the regularity of the
rationally-extended potential $V^{(2)}(x)$ on this domain. The bound-state
wavefunctions of the latter are then given by \cite{cq11b} 
\begin{equation}
\psi^{(2)}_{\nu}(x) = \frac{\eta_l(z)}{g^{(\alpha)}_{\mu}(z)}
y^{(\alpha)}_{\mu+\nu}(z), \qquad \nu=0, 1, 2, \ldots,
\end{equation}
where $y^{(\alpha)}_{\mu+\nu}(z)$ is a polynomial of degree $n=\mu+\nu$
satisfying the second-order differential equation 
\begin{equation}
\left[z \frac{d^2}{dz^2} + \left(\alpha+1-z - 2z \frac{\dot{g}%
^{(\alpha)}_{\mu}}{g^{(\alpha)}_{\mu}}\right) \frac{d}{dz} + (z-\alpha) 
\frac{\dot{g}^{(\alpha)}_{\mu}}{g^{(\alpha)}_{\mu}} + z \frac{\ddot{g}%
^{(\alpha)}_{\mu}}{g^{(\alpha)}_{\mu}}\right] y^{(\alpha)}_{\mu+\nu}(z) = -
\nu y^{(\alpha)}_{\mu+\nu}(z).
\end{equation}

%
%
The $y^{(\alpha)}_{\mu+\nu}$ ($\nu= 0$, 1, 2,~\ldots) polynomials constitute
a complete orthogonal family on the positive half-line with respect to the
positive-definite measure of density 
\begin{equation}
z^{\alpha} e^{-z}/\bigl( g^{(\alpha)}_{\mu}(z)\bigr)^2.
\end{equation}

%
%

\section{\boldmath POLYNOMIAL CHARACTER OF $g_{\protect\mu }^{(\protect%
\alpha )}$ AND BEHAVIOUR NEAR THE ORIGIN}

In \cite{cq11b} it has been conjectured that $g^{(\alpha)}_{\mu}(z)$ is a
polynomial, which can be directly checked in the $k=2$ case, but is not
obvious for higher $k$ values whenever $k-q \ge 2$ (see Eq.~(\ref{eq:g})).
In the following we will give a complete proof of this assertion. In \cite%
{grandati12} the regularity of the extended potential, which is equivalent
to the absence of node of $g^{(\alpha)}_{\mu}$, has been established without
giving the exact highest and lowest degrees coefficients of this polynomial.
We will also determine the value of the lowest-degree coefficient and then
verify explicitly the nodeless character of the Crum Wronskian.

%
%
{}First note that from Eq.~(\ref{eq:wronskn}) we can write for $k-q \ge 2$ 
\begin{equation}
\left\{ 
\begin{array}{c}
\mathcal{W}(\phi_1, \phi_2, \ldots, \phi_k \mid x) = (\omega x)^{k(k-1)/2}
z^{-q(k-q)} (\chi^{\mathrm{I}}_{l^{\prime }})^q (\chi^{\mathrm{II}%
}_{l^{\prime }})^{k-q} g^{(\alpha)}_{\mu}(z), \\[0.2cm] 
\mathcal{W}(\phi_1, \phi_2, \ldots, \phi_{k-1} \mid x) = (\omega
x)^{(k-1)(k-2)/2} z^{-q(k-q-1)} (\chi^{\mathrm{I}}_{l^{\prime }})^q (\chi^{%
\mathrm{II}}_{l^{\prime }})^{k-q-1} g^{(\alpha+1)}_{\mu^{\prime }}(z),%
\end{array}
\right.  \label{eq:wronskg}
\end{equation}
where $l^{\prime }= l+k-2q$ (hence $\alpha^{\prime }= \alpha+k-2q$) and 
\begin{equation}
\left\{ 
\begin{array}{l}
\mu = \sum_{i=1}^k m_i - q(q-1)/2 - (k-q)(k-q-1)/2 + q(k-q), \\[0.2cm] 
\mu^{\prime }= \sum_{i=1}^{k-1} m_i - q(q-1)/2 - (k-q-1)(k-q-2)/2 + q(k-q-1)
\\[0.2cm] 
\hphantom{\mu'} = \mu - m_k + k - 2q - 1.%
\end{array}
\right.
\end{equation}

%
%
Inserting these expressions in Eq.~(\ref{eq:phii}), we arrive at 
\begin{eqnarray}
\phi^{(k)}_{1,2,\ldots,k}(x) & = & (\omega x)^{k-1} z^{-q} \chi^{\mathrm{II}%
}_{l^{\prime }}(z) \frac{g^{(\alpha)}_{\mu}(z)} {g^{(\alpha+1)}_{\mu^{\prime
}}(z)}  \notag \\
& = & (2\omega)^{(k-1)/2} z^{-(2\alpha^{\prime }-2k+4q+1)/4} e^{-z/2} \frac{%
g^{(\alpha)}_{\mu}(z)} {g^{(\alpha+1)}_{\mu^{\prime }}(z)}.  \label{eq:phik}
\end{eqnarray}
%
%
On the other hand, we can obtain a different expression for $%
\phi^{(k)}_{1,2,\ldots,k}$ by using some properties of Wronskians. Indeed,
from Sylvester's theorem \cite{muir}, we can write 
\begin{equation}
\mathcal{W}(\phi_1,\dots,\phi_{k-1},\phi_k \mid x) = \frac{\mathcal{W}\bigl(%
\mathcal{W}(\phi_1,\dots,\phi_{k-2},\phi_{k-1} \mid x), \mathcal{W}%
(\phi_1,\dots,\phi_{k-2},\phi_k \mid x) \mid x\bigr)}{\mathcal{W}%
(\phi_1,\dots,\phi_{k-2} \mid x)}
\end{equation}
and using 
\begin{equation}
\mathcal{W}(f\phi_1,\dots,f\phi_k \mid x) = f^k \mathcal{W}%
(\phi_1,\dots,\phi_k \mid x),
\end{equation}
we have 
\begin{eqnarray}
\lefteqn{{\cal W}(\phi_1,\dots,\phi_{k-1},\phi_k \mid x) = {\cal
W}(\phi_1,\dots,\phi_{k-2} \mid x)}  \notag \\[0.2cm]
&& {} \times \mathcal{W}\left(\frac{\mathcal{W}(\phi_1,\dots,\phi_{k-2},%
\phi_{k-1} \mid x)} {\mathcal{W}(\phi_1,\dots,\phi_{k-2} \mid x)}, \frac{%
\mathcal{W}(\phi_1,\dots,\phi_{k-2},\phi_k \mid x)} {\mathcal{W}%
(\phi_1,\dots,\phi_{k-2} \mid x)}\right. \bigg| x\biggr),
\end{eqnarray}
that is 
\begin{equation}
\phi^{(k)}_{1,2,\ldots,k}(x) = \frac{\mathcal{W}\Bigl(\phi^{(k-1)}_{1,%
\ldots,k-2,k-1},\phi^{(k-1)}_{1,\ldots,k-2,k} \Big| x\Bigr)}{%
\phi^{(k-1)}_{1,\ldots,k-2,k-1}(x)}.  \label{eq:recphi}
\end{equation}

%
%
{}From the Wronskian theorem \cite{messiah}, it follows that the Wronskian
in the numerator of this expression satisfies the property 
\begin{equation}
\Bigl[\mathcal{W}\Bigl(\phi^{(k-1)}_{1,\ldots,k-2,k-1},
\phi^{(k-1)}_{1,\ldots,k-2,k} \Big| \xi\Bigr)\Bigr]^{+\infty}_x =
(E_{k-1}-E_k) \int_x^{+\infty} \phi^{(k-1)}_{1,\ldots,k-2,k-1}(\xi)
\phi^{(k-1)}_{1,\ldots,k-2,k}(\xi) d\xi,  \label{eq:messiah}
\end{equation}
because $\phi^{(k-1)}_{1,\ldots,k-2,k-1}$ and $\phi^{(k-1)}_{1,\ldots,k-2,k}$
correspond to the energies $\sum_{i=1}^{k-1} E_i$ and $\sum_{i=1}^{k-2} E_i
+ E_k$, respectively.

%
%
As a consequence of the exponential factor in Eq.~(\ref{eq:phik}), both $%
\phi^{(k-1)}_{1,\ldots,k-2,k-1}(\xi)$ and $\phi^{(k-1)}_{1,\ldots,k-2,k}(%
\xi) $ vanish for $\xi\to\infty$, which ensures the vanishing of their
Wronskian in the same limit. Hence Equation (\ref{eq:messiah}) reduces to 
\begin{equation}
\mathcal{W}\Bigl(\phi^{(k-1)}_{1,\ldots,k-2,k-1},
\phi^{(k-1)}_{1,\ldots,k-2,k} \Big| x\Bigr) = - (E_{k-1}-E_k)
\int_x^{+\infty} \phi^{(k-1)}_{1,\ldots,k-2,k-1}(\xi)
\phi^{(k-1)}_{1,\ldots,k-2,k}(\xi) d\xi,
\end{equation}
or, on taking Eq.~(\ref{eq:phik}) into account and defining $\zeta = \omega
\xi^2/2$, 
\begin{eqnarray}
\lefteqn{{\cal W}\Bigl(\phi^{(k-1)}_{1,\ldots,k-2,k-1},
\phi^{(k-1)}_{1,\ldots,k-2,k} \Big| x\Bigr) = - (2\omega)^{k-5/2}
(E_{k-1}-E_k)}  \notag \\[0.2cm]
&& {} \times \int_z^{+\infty} \zeta^{k-2q-\alpha^{\prime }-2} e^{-\zeta} 
\frac{g^{(\alpha+1)}_{\mu^{\prime }}(\zeta) g^{(\alpha+1)}_{\bar{\mu}%
^{\prime }}(\zeta)}{\left(g^{(\alpha+2)}_{\mu^{\prime \prime
}}(\zeta)\right)^2} d\zeta,  \label{eq:wronsk}
\end{eqnarray}
where 
\begin{equation}
\left\{ 
\begin{array}{c}
\bar{\mu}^{\prime }= \mu - m_{k-1} + k - 2q -1, \\[0.2cm] 
\mu^{\prime \prime }= \mu - m_{k-1} - m_k + 2(k-2q) - 3.%
\end{array}
\right.
\end{equation}

%
%
Inserting Eqs.~(\ref{eq:phik}) and (\ref{eq:wronsk}) in the right-hand side
of Eq.~(\ref{eq:recphi}), we get the searched for second expression of $%
\phi^{(k)}_{1,2,\ldots,k}(x)$, 
\begin{eqnarray}
\phi^{(k)}_{1,2,\ldots,k}(x) & = & - (2\omega)^{(k-3)/2} (E_{k-1}-E_k) 
\notag \\
&& {} \times \int_z^{+\infty} \zeta^{k-2q-\alpha^{\prime }-2} e^{-\zeta}
g^{(\alpha+1)}_{\mu^{\prime }}(\zeta) g^{(\alpha+1)}_{\bar{\mu}^{\prime
}}(\zeta)/\bigl(g^{(\alpha+2)}_{\mu^{\prime \prime }}(\zeta)\bigr)^2 d\zeta 
\notag \\
&& {} \times \Bigl[z^{(2k-4q-2\alpha^{\prime }-3)/4} e^{-z/2}
g^{(\alpha+1)}_{\mu^{\prime }}(z)/g^{(\alpha+2)}_{\mu^{\prime \prime }}(z) %
\Bigr]^{-1},
\end{eqnarray}
which can also be rewritten as a recursion relation for the $%
g^{(\alpha)}_{\mu}(z)$ functions, 
\begin{eqnarray}
g^{(\alpha)}_{\mu}(z) & = & \frac{E_k-E_{k-1}}{2\omega} z^{-(k-2q-\alpha^{%
\prime }-1)} e^z g^{(\alpha+2)}_{\mu^{\prime \prime }}(z)  \notag \\
&& {} \times \int_z^{+\infty} \zeta^{k-2q-\alpha^{\prime }-2} e^{-\zeta}
g^{(\alpha+1)}_{\mu^{\prime }}(\zeta) g^{(\alpha+1)}_{\bar{\mu}^{\prime
}}(\zeta)/\bigl(g^{(\alpha+2)}_{\mu^{\prime \prime }}(\zeta)\bigr)^2 d\zeta.
\label{eq:recg}
\end{eqnarray}

%
%
On the right-hand side of this relation, $g^{(\alpha+1)}_{\mu^{\prime }}$
and $g^{(\alpha+1)}_{\bar{\mu}^{\prime }}$ correspond to $k^{\prime }=k-1$,
while for $g^{(\alpha+2)}_{\mu^{\prime \prime }}$, $k^{\prime }=k-2$. The
polynomial character of $g^{(\alpha)}_{\mu}(z)$ can be easily obtained by
induction. This is already verified for $k=1$ and $k=2$. Assuming that for $%
k^{\prime }<k$, $g^{(\alpha)}_{\mu}$ is a polynomial with a non-zero
constant term, it turns out that in the neighbourhood of the origin, the
right-hand side of Eq.~(\ref{eq:recg}) has a well-defined value given by 
\begin{equation}
g^{(\alpha)}_{\mu}(z) \simeq \frac{m_k-m_{k-1}}{\alpha^{\prime }-k+2q+1}
g^{(\alpha+1)}_{\mu^{\prime }}(0) g^{(\alpha+1)}_{\bar{\mu}^{\prime }}(0) /
g^{(\alpha+2)}_{\mu^{\prime \prime }}(0) (1 + O(z)),  \label{eq:gatorigin}
\end{equation}
where we have used $E_k = -\omega (\alpha^{\prime }- 2m_k - 1)$.

%
%
{}From Eq.~(\ref{eq:gatorigin}) we immediately deduce that $%
g^{(\alpha)}_{\mu}(z)$ cannot have a pole at the origin and is then a
polynomial with non-zero constant term.

%
%
To completely describe the behaviour of $g^{(\alpha)}_{\mu}(z)$ at the
origin, we need to determine the exact expression of this constant term. For
such a purpose, we will proceed by induction again. Consider first the pure
case I$^k$ ($k=q$), where $\alpha^{\prime }=\alpha-k$. From (\ref{eq:g}) and
(\ref{eq:gamma}), we have 
\begin{equation}
g^{(\alpha)}_{\mu}(z) = 
\begin{vmatrix}
L^{(\alpha^{\prime })}_{m_1}(-z) & \hdots & L^{(\alpha^{\prime })}_{m_k}(-z)
\\ 
\vdots & \ddots & \vdots \\ 
L^{(\alpha^{\prime }+k-1)}_{m_1-k+1}(-z) & \hdots & L^{(\alpha^{\prime
}+k-1)}_{m_k-k+1}(-z)%
\end{vmatrix}%
.
\end{equation}

%
%
On using \cite{erdelyi} 
\begin{equation}
L^{(\alpha)}_m(0) = \frac{(\alpha+1)_m}{m!},
\end{equation}
we easily get 
\begin{eqnarray}
g^{(\alpha)}_{\mu}(0) & = & \frac{(\alpha^{\prime }+k)_{m_1-k+1} \cdots
(\alpha^{\prime }+k)_{m_k-k+1}}{m_1! \cdots m_k!}  \notag \\[0.2cm]
&& {} \times 
\begin{vmatrix}
(\alpha^{\prime }+1) \cdots (\alpha^{\prime }+k-1) & \hdots & 
(\alpha^{\prime }+1) \cdots (\alpha^{\prime }+k-1) \\ 
(\alpha^{\prime }+2) \cdots (\alpha^{\prime }+k-1) m_1 & \hdots & 
(\alpha^{\prime }+2) \cdots (\alpha^{\prime }+k-1) m_k \\ 
\vdots & \ddots & \vdots \\ 
m_1 (m_1-1) \cdots (m_1-k+2) & \hdots & m_k (m_k-1) \cdots (m_k-k+2)%
\end{vmatrix}%
,
\end{eqnarray}
that is (see Eq.~(\ref{eq:vandermonde})) 
\begin{eqnarray}
g^{(\alpha)}_{\mu}(0) & = & \frac{(\alpha^{\prime }+1)_{m_1} \cdots
(\alpha^{\prime }+k)_{m_k-k+1}}{m_1! \cdots m_k!} \Delta(m_1,\ldots,m_k) 
\notag \\
& = & \frac{(\alpha-k+1)_{m_1} \cdots (\alpha)_{m_k-k+1}}{m_1! \cdots m_k!}
\Delta(m_1,\ldots,m_k).
\end{eqnarray}

%
%
Consider now the first mixed case $\mathrm{I}^{k-1} \mathrm{II}$ ($k=q+1$),
where $\alpha^{\prime }=\alpha-k+2$ \cite{cq11b}. We then get 
\begin{equation}
g^{(\alpha)}_{\mu}(z) = 
\begin{vmatrix}
L^{(\alpha^{\prime })}_{m_1}(-z) & \ldots & L^{(\alpha^{\prime
})}_{m_{k-1}}(-z) & z^{k-1} L^{(-\alpha^{\prime })}_{m_k}(z) \\ 
\vdots & \ddots & \vdots & \vdots \\ 
L^{(\alpha^{\prime }+k-2)}_{m_1-k+2}(-z) & \ldots & L^{(\alpha^{\prime
}+k-2)}_{m_{k-1}-k+2}(-z) & (m_k+1)_{k-2} z L^{(-\alpha^{\prime
}-k+2)}_{m_k+k-2}(z) \\ 
L^{(\alpha^{\prime }+k-1)}_{m_1-k+1}(-z) & \ldots & L^{(\alpha^{\prime
}+k-1)}_{m_{k-1}-k+1}(-z) & (m_k+1)_{k-1} L^{(-\alpha^{\prime
}-k+1)}_{m_k+k-1}(z)%
\end{vmatrix}%
,
\end{equation}
which leads to 
\begin{equation}
g^{(\alpha)}_{\mu}(0) = (m_k+1)_{k-1} L^{(-\alpha^{\prime
}-k+1)}_{m_k+k-1}(0) 
\begin{vmatrix}
L^{(\alpha^{\prime })}_{m_1}(0) & \hdots & L^{(\alpha^{\prime
})}_{m_{k-1}}(0) \\ 
\vdots & \ddots & \vdots \\ 
L^{(\alpha^{\prime }+k-2)}_{m_1-k+2}(0) & \hdots & L^{(\alpha^{\prime
}+k-2)}_{m_{k-1}-k+2}(0)%
\end{vmatrix}%
.
\end{equation}

%
%
The right-hand member determinant is exactly the constant term of the $g$
polynomial in the pure case I$^{k-1}$ and we deduce from the previous result
that 
\begin{eqnarray}
g^{(\alpha)}_{\mu}(0) & = & \frac{(\alpha^{\prime }+1)_{m_1} \cdots
(\alpha^{\prime }+k-1)_{m_{k-1}-k+2}}{m_1! \cdots m_{k-1}!}
\Delta(m_1,\ldots,m_{k-1})  \notag \\
&& {}\times (m_k+1)_{k-1} (-1)^{m_k+k-1} \frac{(\alpha^{\prime
}-m_k)_{m_k+k-1}}{(m_k+k-1)!},
\end{eqnarray}
or 
\begin{eqnarray}
g^{(\alpha)}_{\mu}(0) & = & (-1)^{m_k+k-1} \frac{(\alpha^{\prime }+1)_{m_1}
\cdots (\alpha^{\prime }+k-1)_{m_{k-1}-k+2} (\alpha^{\prime }-m_k)_{m_k+k-1}%
}{m_1! \cdots m_k!}  \notag \\
&& {}\times \Delta(m_1,\ldots,m_{k-1})  \notag \\
& = & (-1)^{m_k+k-1} \frac{(\alpha-k+3)_{m_1} \cdots
(\alpha+1)_{m_{k-1}-k+2} (\alpha-k+2-m_k)_{m_k+k-1}}{m_1! \cdots m_k!} 
\notag \\
&& {}\times \Delta(m_1,\ldots,m_{k-1}).
\end{eqnarray}

%
%
{}For the general mixed case $\mathrm{I}^q \mathrm{II}^{k-q}$, we now make
the following hypothesis 
\begin{eqnarray}
g^{(\alpha)}_{\mu}(0) & = & (-1)^{\sigma} \frac{\Delta(m_1,\ldots,m_q)
\Delta(m_{q+1},\ldots,m_k)}{m_1! \cdots m_k!}  \notag \\
&& {} \times \left((\alpha+k-2q+1)_{m_1} \cdots (\alpha+k-q)_{m_q-q+1}\right)
\notag \\
&& {} \times \left((\alpha+k-2q-m_{q+1})_{m_{q+1}+q} \cdots
(\alpha+k-2q-m_k)_{m_k+2q-k+1}\right),  \label{eq:g0}
\end{eqnarray}
with $\sigma$ defined in Eq.~(\ref{eq:musigma}), and prove it by induction
over $k$.

%
%
This conjecture is verified in the $k=1$ and $k=2$ cases. Since it is also
valid for the two previous examples ($k-q=0$ or 1), it only remains to prove
it whenever $k-q \ge 2$. In such cases, the last two seed functions $%
\phi_{k-1}$ and $\phi_k$, used in $k$th-order SUSYQM, are necessarily of
type II. Suppose that Eq.~(\ref{eq:g0}) is verified for $k^{\prime }<k$. We
may therefore write 
\begin{eqnarray}
&& g^{(\alpha+2)}_{\mu^{\prime \prime }}(0) = (-1)^{\sigma^{\prime \prime }} 
\frac{\Delta(m_1,\ldots,m_q) \Delta(m_{q+1},\ldots,m_{k-2})} {m_1! \cdots
m_{k-2}!}  \notag \\
&& \quad {} \times \left((\alpha+k-2q+1)_{m_1} \cdots
(\alpha+k-q)_{m_q-q+1}\right)  \notag \\
&& \quad {} \times \left((\alpha+k-2q-m_{q+1})_{m_{q+1}+q} \cdots
(\alpha+k-2q-m_{k-2})_{m_{k-2}+2q-k+3}\right),
\end{eqnarray}
\begin{eqnarray}
&& g^{(\alpha+1)}_{\mu^{\prime }}(0) = (-1)^{\sigma^{\prime }} \frac{%
\Delta(m_1,\ldots,m_q) \Delta(m_{q+1},\ldots,m_{k-1})} {m_1! \cdots m_{k-1}!}
\notag \\
&& \quad {} \times \left((\alpha+k-2q+1)_{m_1} \cdots
(\alpha+k-q)_{m_q-q+1}\right)  \notag \\
&& \quad {} \times \left((\alpha+k-2q-m_{q+1})_{m_{q+1}+q} \cdots
(\alpha+k-2q-m_{k-1})_{m_{k-1}+2q-k+2}\right),
\end{eqnarray}
and 
\begin{eqnarray}
&& g^{(\alpha+1)}_{\bar{\mu}^{\prime }}(0) = (-1)^{\bar{\sigma}^{\prime }} 
\frac{\Delta(m_1,\ldots,m_q) \Delta(m_{q+1},\ldots,m_{k-2},m_k)}{m_1! \cdots
m_{k-2}!m_k!}  \notag \\
&& \quad {} \times \left((\alpha+k-2q+1)_{m_1} \cdots
(\alpha+k-q)_{m_q-q+1}\right)  \notag \\
&& \quad {} \times \left((\alpha+k-2q-m_{q+1})_{m_{q+1}+q} \cdots
(\alpha+k-2q-m_{k-2})_{m_{k-2}+2q-k+3} \right),  \notag \\
&& \quad {} \times (\alpha+k-2q-m_k)_{m_k+2q-k+2},
\end{eqnarray}
with 
\begin{equation}
\left\{ 
\begin{array}{c}
\sigma^{\prime }= \sum_{l=q+1}^{k-1} m_l + q(k-q-1), \\[0.2cm] 
\sigma^{\prime \prime }= \sum_{l=q+1}^{k-2} m_l + q(k-q-2), \\[0.2cm] 
\bar{\sigma}^{\prime }= \sum_{l=q+1}^{k-2} m_l + m_k + q(k-q-1).%
\end{array}
\right.
\end{equation}

%
%
{}From Eq.~(\ref{eq:gatorigin}), we obtain the recurrence relation 
\begin{equation}
g^{(\alpha)}_{\mu}(0) = \frac{m_k - m_{k-1}}{\alpha+1} g^{(\alpha+1)}_{\mu^{%
\prime }}(0) g^{(\alpha+1)}_{\bar{\mu}^{\prime }}(0) /
g^{(\alpha+2)}_{\mu^{\prime \prime }}(0),  \label{eq:recg0}
\end{equation}
which combined with the previous expressions yields 
\begin{eqnarray}
g^{(\alpha)}_{\mu}(0) & = & \frac{(-1)^{\sigma^{\prime }+\bar{\sigma}%
^{\prime }-\sigma^{\prime \prime }} (m_k-m_{k-1})}{(\alpha+1) m_1! \cdots
m_k!}  \notag \\
&& {} \times \frac{\Delta(m_1,\ldots,m_q) \Delta(m_{q+1},\ldots,m_{k-1})
\Delta(m_{q+1},\ldots,m_{k-2},m_k)} {\Delta(m_{q+1},\ldots,m_{k-2})}  \notag
\\
&& {} \times \left((\alpha+k-2q+1)_{m_1} \cdots (\alpha+k-q)_{m_q-q+1}\right)
\notag \\
&& {} \times \left((\alpha+k-2q-m_{q+1})_{m_{q+1}+q} \cdots
(\alpha+k-2q-m_{k-2})_{m_{k-2}+2q-k+3}\right)  \notag \\
&& {} \times (\alpha+k-2q-m_{k-1})_{m_{k-1}+2q-k+2}
(\alpha+k-2q-m_k)_{m_k+2q-k+2}.
\end{eqnarray}

%
%
Here we note that 
\begin{equation}
\sigma^{\prime }+ \bar{\sigma}^{\prime }- \sigma^{\prime \prime }=
\sum_{l=q+1}^{k-2} m_l + m_{k-1} + m_k + q(k-q) = \sigma,
\end{equation}
\begin{equation}
\frac{(\alpha+k-2q-m_k)_{m_k+2q-k+2}}{\alpha+1} =
(\alpha+k-2q-m_k)_{m_k+2q-k+1},
\end{equation}
and 
\begin{eqnarray}
\lefteqn{(m_k-m_{k-1}) \frac{\Delta(m_{q+1},\ldots,m_{k-1})
\Delta(m_{q+1},\ldots,m_{k-2},m_k)} {\Delta(m_{q+1},\ldots,m_{k-2})}}  \notag
\\
&& = (m_k-m_{k-1}) \left(\prod_{i=q+1}^{k-2} \prod_{j=i+1}^{k-1}
(m_j-m_i)\right) \left(\prod_{i=q+1}^{k-2} (m_k-m_i)\right)  \notag \\
&& = \prod_{i=q+1}^{k-1} \prod_{j=i+1}^k (m_j-m_i) =
\Delta(m_{q+1},\ldots,m_k).
\end{eqnarray}
The final result for $g^{(\alpha)}_{\mu}(0)$ is therefore given by Eq.~(\ref%
{eq:g0}), showing that the conjecture is still verified for $k$, which ends
its proof.

%
%
Let us observe that in terms of $\alpha^{\prime }$, Equation (\ref{eq:g0})
can be rewritten as 
\begin{eqnarray}
g^{(\alpha)}_{\mu}(0) & = & (-1)^{\sigma} \frac{\Delta(m_1,\ldots,m_q)
\Delta(m_{q+1},\ldots,m_k)}{m_1! \cdots m_k!}  \notag \\
&& {} \times \left((\alpha^{\prime }+1)_{m_1} \cdots (\alpha^{\prime
}+q)_{m_q-q+1}\right)  \notag \\
&& {} \times \left((\alpha^{\prime }-m_{q+1})_{m_{q+1}+q} \cdots
(\alpha^{\prime }-m_k)_{m_k+2q-k+1}\right).  \label{eq:g0bis}
\end{eqnarray}
Since the angular momentum $l^{\prime }$ in the starting potential $%
V^{(1)}(x)$ is necessarily nonnegative, the inequality $\alpha^{\prime }>
-1/2$ is fulfilled. Furthermore, we have assumed $0 < m_1 < \cdots <m_q$ and 
$0 < m_{q+1} < \cdots < m_k < \alpha^{\prime }$. As a consequence, the two
Vandermonde determinants in Eq.~(\ref{eq:g0bis}) are positive and 
\begin{equation}
(\alpha^{\prime }+1)_{m_1} \cdots (\alpha^{\prime }+q)_{m_q-q+1}
(\alpha^{\prime }-m_{q+1})_{m_{q+1}+q} \cdots (\alpha^{\prime
}-m_k)_{m_k+2q-k+1} > 0.
\end{equation}
We conclude that the sign of $g^{(\alpha)}_{\mu}(0)$ is given by $%
(-1)^{\sigma}$, hence is identical with the one of $\mathcal{C}%
^{(\alpha)}_{\mu}$ (see Eq.~(\ref{eq:musigma})). The combination of this
result with the disconjugacy theorem of Sec.~II ensures the regularity of
the extended potential $V^{(2)}(x)$.

%
%

\section{CONCLUSION}

In the present work, we have established some further properties of the
function $g^{(\alpha)}_{\mu}(z)$ appearing in the denominator of
rationally-extended isotonic oscillator (or radial oscillator) potentials
constructed in $k$th-order SUSYQM \cite{cq11b} or using the multistep
Darboux-B\"acklund transformation method \cite{grandati12}. We have first
confirmed that $g^{(\alpha)}_{\mu}(z)$ cannot have any pole at the origin
and is a polynomial with non-zero constant term, a property that was
conjectured in Ref.~\cite{cq11b}. We have then determined the value of this
constant term by induction over $k$ and shown that it has the same sign as
the coefficient of the previously determined highest-degree term $z^{\mu}$ 
\cite{cq11b}.

%
%
These results have been combined with the disconjugacy theorem applied to
Schr\"odinger equation \cite{grandati12} to prove the absence of zeros of $%
g^{(\alpha)}_{\mu}(z)$ on the positive half-line and, therefore, the
regularity of the corresponding rationally-extended potential. This has
provided a demonstration of the latter property, which is more explicit than
that already given in Ref.~\cite{grandati12}.

%
%
Our work has illustrated the power of the disconjugacy properties of
second-order differential equations of Schr\"odinger type to check the
regularity of rational extensions of well-known quantum potentials. Other
examples of application might be found in cases where no direct information
on the denominator zeros is available. This will be the object of further
investigations.

%
%

\end{document}